# Trajectories of the $S$-matrix poles in Salamon-Vertse potential

A. Rácz[1], P. Salamon[1,2] and T. Vertse[1,2]

[1]*University of Debrecen, Faculty of Informatics,*

*PO Box 12, H–4010 Debrecen, Hungary*

[2]*Institute of Nuclear Research of the Hungarian Academy of Sciences,*

*Debrecen, PO Box 51, H–4001, Hungary*

## Abstract

The trajectories of $S$-matrix poles are calculated in the finite-range phenomenological potential introduced recently by P. Salamon and T. Vertse [1] (SV). The trajectories of the resonance poles in this SV potential are compared to the corresponding trajectories in a cut-off Woods-Saxon (WS) potential for $l > 0$. The dependence on the cut-off radius is demonstrated. The starting points of the trajectories turn out to be related to the average ranges of the two terms in the SV potential.

PACS numbers: 21.10.Pc,21.30.-x,21.60.Cs

Keywords: $S$-matrix

Solutions, of the Schrödinger equations, that belong to poles of the scattering matrix are often used in reaction calculations and also in the description of weakly bound states. Certain pole solutions can be used as single particle basis functions in a so called Berggren representation [2], and they are useful in the theoretical description of slightly bound or unbound states in the framework of the Gamow shell model [3] (GSM). The GSM is a powerful tool for the description of the new nuclei produced by the radioactive beam facilities recently. For the numerical calculation of the $S$-matrix poles in the Woods-Saxon (WS) potential a new method was suggested recently [4].

The WS potential form is often used in reaction calculations. Although the WS potential is considered a finite-range potential, its value becomes zero only at infinite distances. The radial Schrödinger equation with this potential has analytical solutions [5] only for $l = 0$. Therefore, we have to calculate its solutions (eigenvalues and eigenfunctions) by using numerical methods for the integration. The typical way of the solution is that we integrate the radial equation numerically in an interval $r \in [0, R_{match}]$ of the radial distance $r$ where the nuclear potential is large, and we match the calculated solution at a distance $r = R_{match}$ Ref. [9], or at two different distances in Ref. [4] to the solution of the asymptotic differential equation, which is the Coulomb or the Ricatti-Hankel differential equation. Thus, in this solution we effectively use a modified version of the WS potential in which the potential is cut to zero at a finite distance $R_{max}$ artificially. The cut-off WS potential is the following:

$$V^{WS}(r) = V_0 f^{WS}(r) \; , \tag{1}$$

where the radial shape is

$$f^{WS}(r) = - \begin{cases} \frac{1}{1+e^{\frac{r-R}{a}}} & , \text{if } r < R_{max} \\ 0 & , \text{if } r \geq R_{max} \; . \end{cases} \tag{2}$$

The solution and, consequently, the resulting pole position does depend on the value of the cut-off radius $R_{max}$. Therefore, the cut-off radius $R_{max}$ is a parameter of the WS form in Eq.(2), in addition to its depth $V_0$, radius $R$ and diffuseness $a$. In publications, however, the value of $R_{max}$ is very seldom specified because most physicists think that the cut has negligible effect on the calculated quantities if $R_{max}$ is large and thus the jump at the cut-off is quite small.

In a recent work by P. Salamon and T. Vertse [1] it was discussed that the positions of the broad resonances might be very sensitive to the $R_{max}$ value and the authors proposed an alternative to the cut-off WS potential. The advantage of the proposed new potential (SV) is that its range is finite in the strict sense since it is exactly zero at a value $\rho_{max}$. From mathematical point of view the SV potential is nicer than the cut-off WS potential since its derivative exists everywhere in contrast to the latter. Indeed the shape of the SV potential becomes zero smoothly, and even its derivatives become zero at and beyond $r = \rho_{max}$.

The objective of this Brief Report is to show the implications of the tail of the WS potential for the positions of the $S$-matrix poles, and their remedy achieved by the SV potential.

The sensitivity to $R_{max}$ can be seen if we follow the position of the $S$-matrix poles in the complex wave number plane ($k$-plane) as the strength of the potential $V_0$ is changed. These curves are the so called pole trajectories. The shapes of the pole trajectories do depend on the radial shape of the potential concerned. The pole trajectories in a square well/barrier potential had been studied by Nussenzweig [6] long time ago. In a square well the analytical form of the solution is known inside the potential radius $R_{sq}$ as well, and the pole positions are obtained as roots of a simple transcendental equation.

Certain features of the pole trajectories in the square well remain valid for realistic (diffuse) potentials if their range is finite. Other features however do depend on the radial shape. The pole trajectories of the cut-off WS form have not been studied so extensively as those of the square well. In a work by J. Bang et al. [7] the trajectories of the resonant poles were given for a WS potential cut off at a fixed $R_{max}$ value, but the $R_{max}$ dependence of the trajectories was not explored.

We have now calculate the pole trajectories in cut-off WS potentials with different cut-off radii, and compare them to the similar trajectories in the SV potentials. We present the calculations for $l > 0$ only since in the $l = 0$ case we have no centrifugal barrier to produce narrow resonances, and therefore this case is of reduced importance.

The radial shape of the potentials is fixed, and only the strength is varied along a pole trajectory. For the sake of comparison, we rewrite the original SV potential in Ref. [1] into the form

$$V^{SV}(r) = V_0 f^{SV}(r) , \qquad (3)$$

where $f^{SV}(r)$ is taken a linear combination

$$f^{SV}(r, c_1, \rho_0, \rho_1) = f_{\rho_0}(r) + c_1 f'_{\rho_1}(r) . \qquad (4)$$

Where

$$f_\rho(r) = \begin{cases} -e^{\frac{r^2}{r^2-\rho^2}} & , \text{if } r < \rho \\ 0 & , \text{if } r \geq \rho \,, \end{cases} \quad (5)$$

and $f'_\rho(r)$ is its derivative:

$$f'_\rho(r) = \begin{cases} -\frac{2r\rho^2}{(r^2-\rho^2)^2}\, e^{\frac{r^2}{r^2-\rho^2}} & , \text{if } r < \rho \\ 0 & , \text{if } r \geq \rho \,. \end{cases} \quad (6)$$

Since both terms in Eq.(4) have finite ranges, the SV potential of Eq.(3) has a finite-range of $\rho_{max} = max\{\rho_0, \rho_1\}$, i.e., $f^{SV}(r \geq \rho_{max}) = 0$. With the usual choices ($\rho_0 > \rho_1$), the range is $\rho_{max} = \rho_0$.

In order to make the shape of the SV potential similar to the $f^{WS}(r)$ shape, we adjust the range parameters $\rho_0$, $\rho_1$ and the weight $c_1$ so as to minimize the integral of the squared differences:

$$\Delta(\rho_0, \rho_1, c_1) = \int_0^{\rho_{max}} [f^{SV}(r, c_1, \rho_0, \rho_1) - f^{WS}(r)]^2 dr \,. \quad (7)$$

In general, the $R_{max}$ value is large ($\rho_{max} < R_{max}$). The fitting fixes the geometrical parameters of the SV potential ($\rho_0, \rho_1, c_1$), so that $f^{SV}(r)$ depends only on $R$ and $a$ of the WS shape $f^{WS}(r)$ and does not depend on $R_{max}$. The quality of the fit can be characterized by the minimum $\Delta_{min}$ of $\Delta$.

We consider potentials whose shapes are suitable for describing $^{16}$O and $^{208}$Pb. We compare the radial shapes of the best fit SV potentials and the original WS potentials in Fig. 1. One can see that the shapes of the WS forms are reproduced reasonably well both inside and in the surface regions. There are, however, considerable differences in the tail region, where the SV shape goes to zero faster than the WS shape.

In Table I we summarize the geometrical parameters of the WS and SV potentials used in this work. The parameters of the WS potential for oxygen and for lead were taken from J. Bang et al. [7] and from P. Curutchet et al. [8], respectively. The spin-orbit terms were neglected for the sake of simplicity. The SV potential parameters in Table I are the results of the minimization of the deviation $\Delta$ in Eq.(7). The values of $\Delta_{min}$ are also given in the Table.

We can notice that the shape of the SV potential in Eq.(3) has three parameters ($c_1$, $\rho_0$, $\rho_1$) just as the WS shape ($R$, $a$, $R_{max}$), but the SV potential has a natural finite-range: $\rho_{max}$.

We can observe in Table I that the average values of the two ranges $\bar{\rho}$ show some similarity to the radius parameter $R$ of the WS potential but $\bar{\rho}$ is larger in both cases.

We shall only present trajectories for the decaying resonances since the trajectory of a capturing resonance is simply the mirror image of the trajectory of a decaying resonance.

The pole energies and radial wave functions have been calculated by a modified version of the computer code GAMOW [9]. The node number $n$ starts with $n = 1$, by convention. This $n$ is the actual radial node number in the limit of large potential depth when the pole becomes a bound-state pole.

For conveniences we start the trajectories with a small $V_0$ at a small positive strength, $V_0 = 5$ keV. We denote the wave number of the pole belonging to $V_0 = 5$ keV by $k^0_{n,l}$. After finding $k^0_{n,l}$ we increment the $V_0$ value by small steps repeatedly until the resonance pole reaches the imaginary $k$-axis. For a high $l$ value or for a large node number the imaginary $k$-axis is reached at a very large $V_0$ value, which is much deeper than the realistic potential well.

Nussenzweig [6] found that for a square well $l = 0$ the pole trajectories approach the $Re(k_n) = \frac{n\pi}{R_{sq}}$ lines with the depth approaching to zero. For $l = 1$ the starting values of the pole trajectories approach the $Re(k_n) = \frac{(2n-1)\pi}{2R_{sq}}$ values.

Bang et al. [7] calculated pole trajectories in a cut-off WS potential for $^{16}$O with a cut-off radius $R_{max} = 9.4$ fm for $l = 0$ and $l = 1$. Figs. 5 and 6 of that work suggest that the density of the $l = 0$ and $l = 1$ poles are close to the densities of the square well poles belonging to $R_{max} = R_{sq}$, where $R_{sq}$ denotes the radius of the square well. But the shapes of the cut-off WS trajectories were quite different from the shapes of the square well trajectories.

In Fig. 2 we present the trajectories of the $n = 2$, $l = 1$ resonance pole for $^{16}$O for WS potentials cut off at different distances $R_{max}$. The curve with $R_{max} = 9.4$ fm resembles the shape of the $n = 2$ curve in Fig. 6 of Ref. [7]. It should not be the same since we neglected the spin-orbit term of the potential. We can see that for different $R_{max}$ values the real parts of the starting points $k^0_{2,1}$ are different. The larger the $R_{max}$ value, the smaller $Re(k^0_{2,1})$ is obtained.

We define a sort of *pole density* at the starting points belonging to $V_0 = 5$ keV as the number of $Re(k^0_{n,l})$ values of the same $l$, falling into a $Re(k)$ interval of unit length. Thus, the smaller is the distance between the starting values with node number $n$ and $n + 1$, the higher is the pole density in a given partial wave $l$. The density of the resonance poles in WS potential increases with increasing cut-off radii. This is a typical feature of the trajectories calculated in cut-off WS potentials for all $l$ and $n$ values.

Now we turn to studying trajectories in SV potential. In Fig. 3 one can see the trajectories of the $l = 1$ resonance pole for the SV potential for different $n$ values. The vertical lines in the figure denote the $k$-values for which $Re(k) = \frac{(2n-1)\pi}{2\bar{\rho}}$. The starting points of the trajectories with different $n$ values are close to the corresponding vertical lines of the figure. Therefore, we conclude that for the SV potential the average value of the ranges of the two terms play the same role as the cut-off radius $R_{max}$ of the WS potential. This rule is also applies to the $l = 2$ case, which is shown in Fig. 4. However, in this case $l$ is even, therefore the vertical lines correspond to the $k$ values with $Re(k) = \frac{n\pi}{\bar{\rho}}$. For $l = 0$ the real parts of the starting $k^0_{0,n}$ values are similar to those of the $l = 2$ $k$-trajectories ($Re(k^0_{2,n})$). The shapes of the trajectories are however different, because for $l = 0$, the decaying and capturing resonance poles meet at the imaginary $k$-axis well below the origin.

For $^{208}$Pb the pole trajectories in the WS potential depend on the $R_{max}$ value in the same way as observed for $^{16}$O. The larger the $R_{max}$ value, the smaller is the abscissa of the starting point of the trajectory.

For the SV potential the pole trajectories are very similar to the those for $^{16}$O, therefore, we only show the $l = 2$ case with different $n$ values.

In Fig. 5 we show the resonance pole trajectories for $l = 2$ and for $n = 1, 2, 3, 4$ values. The curves are similar to the ones in Fig. 4. The starting points of the trajectories $k^0_{n,2}$ are close to the values denoted by vertical lines at $Re(k) = \frac{n\pi}{\bar{\rho}}$ with $\bar{\rho} = 9.644$ fm.

To sum up, a common feature of the trajectories in the three finite-range potentials (square well, cut-off WS and SV types) is that the density of the resonant poles is influenced by the range of the potentials. The rule derived by Nussenzweig [6] for square well is that for $V_0 \to 0$ the the distance of the consecutive resonance poles $Re(k_{n+1,l} - k_{n,l})$ tend to $\frac{\pi}{R_{sq}}$ for any $l$ value. Bang et. al. [7] conjectured that this rule applies for cut-off WS if we replace the $R_{sq}$ by the cut-off radius $R_{max}$. They also conjectured that for $V_0 \to 0$ the $l = 0$ poles approach the $Re(k_{n,0}) = \frac{n\pi}{R_{max}}$ and the $l = 1$ poles approach the $Re(k_{n,1}) = \frac{(2n-1)\pi}{2R_{max}}$. Our results in Fig. 2 support these conjecture qualitatively.

In the SV potential case the shapes of the trajectories are different from that of the WS potential and the starting $k$-value depends on the average of the ranges of the two terms, rather than the range, i.e., on the value of $\rho_{max}$. Since $\bar{\rho}$ is close to the radius $R$ of the WS form, the distance between the starting $Re(k)$-values for consecutive $n$ values is governed by the radius of the WS potential to which the parameters of the SV potential were fitted, and not by the cut-off radius $R_{max}$.

When we calculate the pole energies numerically, we have to take the matching radius $R_{match}$, or both values of the matching distances used in Ref. [4] larger than the ranges of the potentials ($R_{max}$ and $\rho_{max}$).

The dependence of the trajectory on the $R_{max}$ value points to the importance of specifying the value of the $R_{max}$ used in the calculations in order to avoid uncertainties in the pole positions. The uncertainty is appreciable for broad resonances.

Contrary to common belief, this strong $R_{max}$ dependence in the cut-off WS potential cannot be diminished by increasing the value of the cut-off radius.

The authors are grateful to R. G. Lovas for valuable discussions. This work was supported by the TÁMOP 4.2.1./B-09/1/KONV-2010-0007/IK/IT project. The project is implemented through the New Hungary Development Plan co-financed by the European Social Fund, and the European Regional Development Fund.

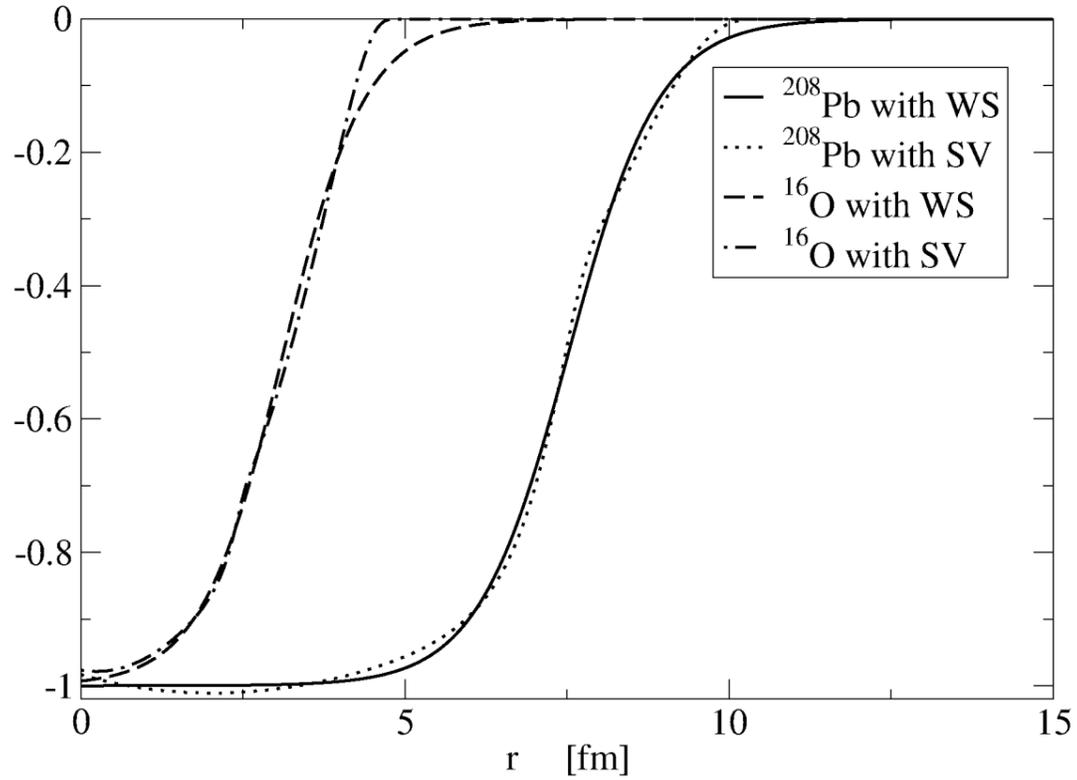

FIG. 1: Comparison of the radial shapes $f^{WS}(r)$ in Eq.(2) and $f^{SV}(r)$ in Eq.(3) for the $^{16}$O and for the $^{208}$Pb nuclei with parameters listed in Table I.

| A | $c_1$ | $\rho_0$ | $\rho_1$ | $\bar{\rho}$ | $R$ | $a$ | $\Delta_{min}$ |
|---|---|---|---|---|---|---|---|
| 16 | 0.066 | 5.082 | 2.707 | 3.892 | 3.125 | 0.63 | $5.6 \times 10^{-3}$ |
| 208 | 0.997 | 10.963 | 8.328 | 9.644 | 7.525 | 0.7 | $2.3 \times 10^{-3}$ |

TABLE I: Geometrical parameters of the SV and WS potentials and the minimal deviation for nuclei $^{16}$O and $^{208}$Pb. All distances are in fm units.

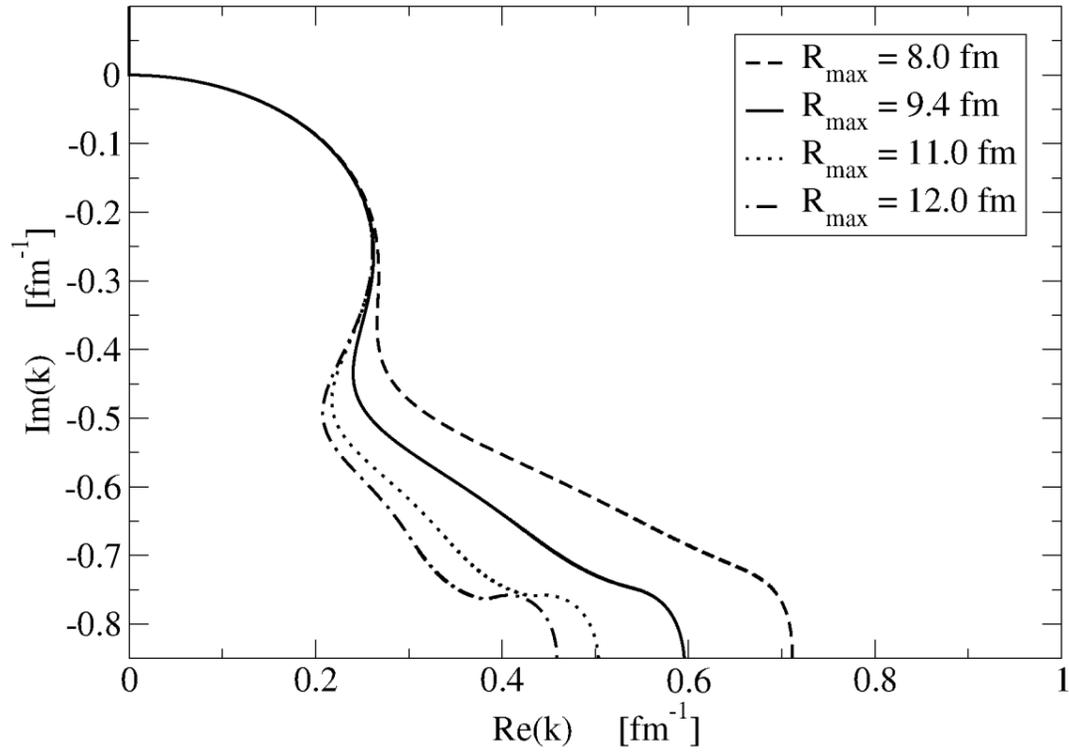

FIG. 2: Trajectories of the $n = 2$ resonant poles in the complex $k$-plane for $l = 1$ for WS potentials for $^{16}$O with different values of the cut-off radius $R_{max}$. The $Re(k) = \frac{3\pi}{2R_{max}}$ values are: 0.59 fm, 0.5 fm, 0.43 fm and 0.39 fm, for increasing $R_{max}$ values, respectively.

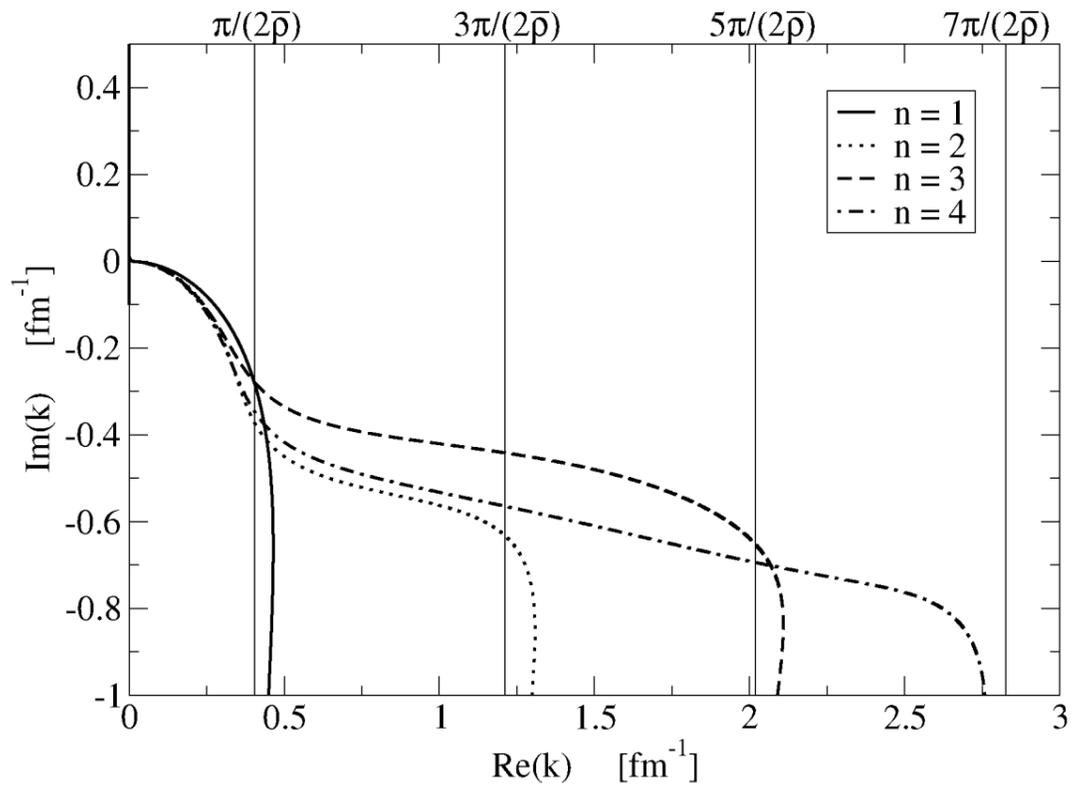

FIG. 3: Trajectories of the resonant poles in the complex $k$-plane for $l = 1$ for SV potentials for different $n$ values for $^{16}$O.

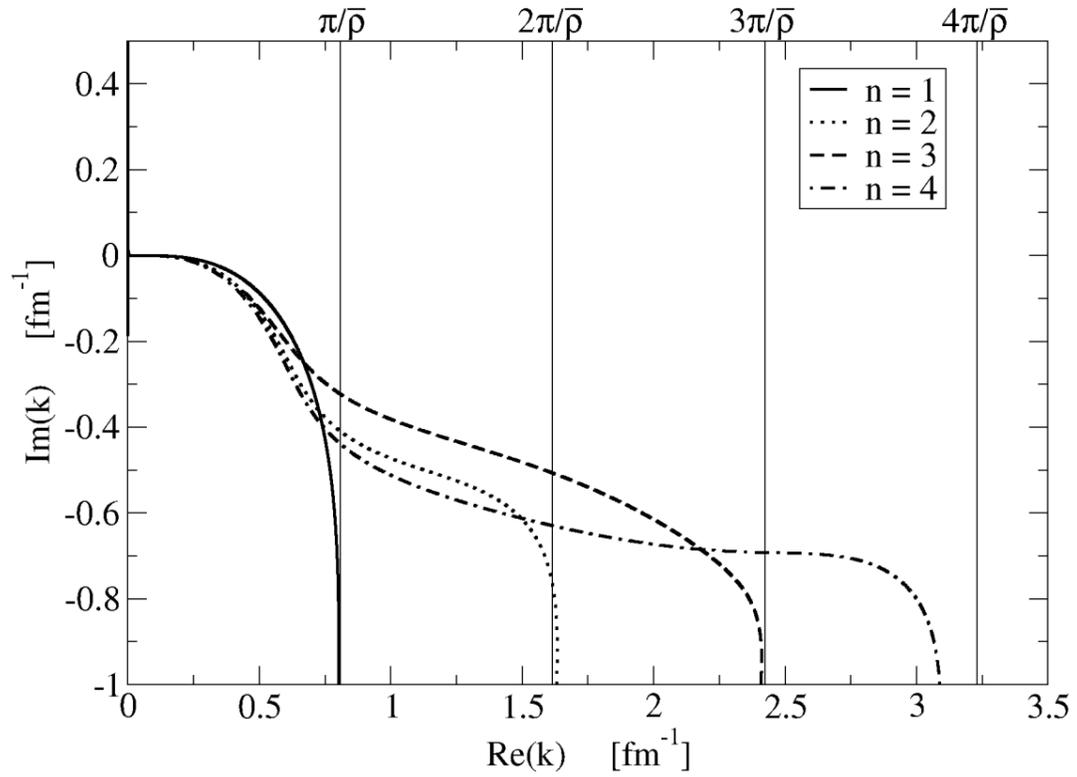

FIG. 4: Trajectories of the resonant poles in the complex $k$-plane for $l = 2$ for SV potentials for different $n$ values for $^{16}$O.

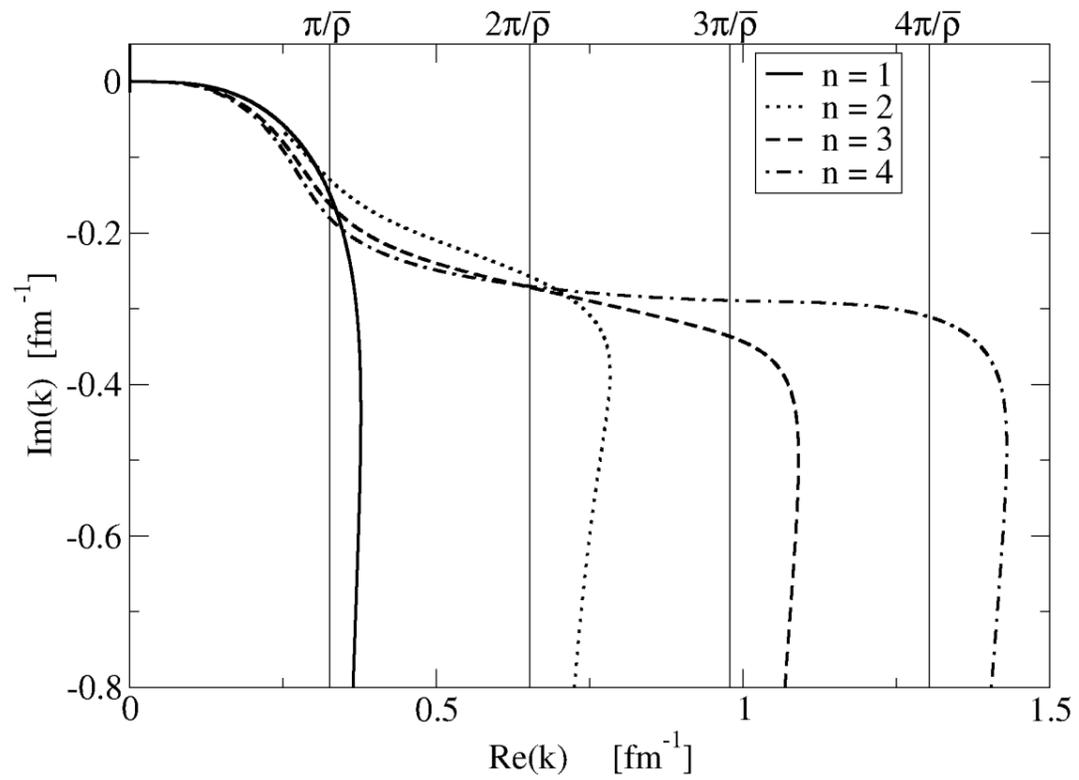

FIG. 5: Trajectories of the resonant poles in the complex $k$-plane in SV potential for $l = 2$ for different $n$ values for $^{208}$Pb.